\documentclass[preprint,11pt,fleqn]{elsarticle}

\usepackage{amsthm} % theorems AMS style
\usepackage{amssymb} % math symbols
\usepackage{amstext} % typeset text in math environments
\usepackage{amsmath} % AMS math­e­mat­i­cal fa­cil­i­ties
\usepackage{bm} % bold symbols in math
\usepackage{graphicx} % enhanced support for graphics
\usepackage{epstopdf} % con­vert EPS to 'en­cap­su­lated' PDF us­ing ghostscript
\usepackage{subcaption} % support for subcaptions
\usepackage{float}
\usepackage{tikz}
\usepackage{lineno,hyperref}
\usepackage[capitalize]{cleveref} % in­tel­li­gent cross-ref­er­enc­ing
\usepackage[makeroom]{cancel} %cancel terms
\usepackage{multirow} 
\usepackage{booktabs} 
\usepackage{lscape}
\usepackage{tabularx} % in the preamble
\usepackage{adjustbox}
\usepackage{tablefootnote}
\usepackage{threeparttablex}
\usepackage{longtable}

\makeatletter
\newcommand{\mathleft}{\@fleqntrue\@mathmargin0pt}
\newcommand{\mathcenter}{\@fleqnfalse}
\makeatother

\modulolinenumbers[5]

\crefalias{subequation}{equation} % subequations counter = equations counter
\crefalias{eqnarray}{equation} % subequations counter = equations counter
\crefformat{pluraleq}{Eqs.~(#2#1#3)} % defining "plural" equations  
\Crefformat{pluraleq}{Equations~(#2#1#3)} % defining "plural" equations  

\def\bal#1\nal{\begin{align}#1\end{align}}
\def\bala#1\nala{\begin{align*}#1\end{align*}}
\def\bsub#1\nsub{\begin{subequations}#1\end{subequations}}

\journal{ }

 \biboptions{sort&compress}

\begin{document}

\begin{frontmatter}

\title{An improved spectral approach for solving the nonclassical neutron particle transport equation
}\author[iprj]{L.R.C. Moraes\corref{cor1}}
\author[umc]{J.K.~Patel\fnref{patel}}
\author[iprj]{R. C. Barros\fnref{barros}}
\author[osu]{R.~Vasques\fnref{vasques}}

\address[iprj]{Universidade do Estado do Rio de Janeiro, Departamento de Modelagem Computacional – IPRJ, Rua Bonfim 25, 28625-570, Nova Friburgo, RJ, Brazil}
\address[umc]{University of Michigan, Department of Nuclear Engineering and Radiological Sciences,\\ 2355 Bonisteel Blvd, Ann Arbor, MI 48109}
\address[osu]{The Ohio State University, Department of Mechanical and Aerospace Engineering,\\ 201 W. 19th Avenue, Columbus, OH 43210}

\cortext[cor1]{Corresponding author:  lrcmoraes@iprj.uerj.br}
\fntext[patel]{jakpatel@umich.edu}
%\fntext[patel]{patel.3545@osu.edu}
\fntext[barros]{ricardob@iprj.uerj.br}
\fntext[vasques]{richard.vasques@fulbrightmail.org}

\begin{abstract}
An improvement modification of the Spectral Approach (SA) used for approximating the nonclassical neutral particle transport equation is described in this work. The main focus of the modified SA lies on a slight modification of the nonclassical angular flux representation as a function of truncated Laguerre series. This leads, in some cases, to a considerable decrease of the Laguerre truncation order required to generate accurate solutions. Numerical results are given to illustrate this proposed improvement.
\end{abstract}

\end{frontmatter}
%\vspace*{-0cm}
\section{Introduction}\label{sec1}
The nonclassical neutral particle transport is a branch of transport theory focused on problems where the particle's distance-to-collision is not exponentially distributed. Larsen in the literature \cite{Larsen01} first derived an equation to mathematically model these problems. This equation for one-speed calculations with isotropic scattering appears as
\begin{subequations}
\begin{flalign}\label{e1.1a}
\begin{aligned}
&\frac{\partial}{\partial s}\Psi(\boldsymbol{x},\boldsymbol{\Omega},s) + \boldsymbol{\Omega}\boldsymbol{\cdot}\nabla\Psi(\boldsymbol{x},\boldsymbol{\Omega},s)+ \Sigma_{t}(\boldsymbol{\Omega},s)\Psi(\boldsymbol{x},\boldsymbol{\Omega},s) =\\& \frac{\delta(s)}{4\pi}\left[c\int_{4\pi}^{}\int_{0}^{\infty}\Sigma_{t}(\boldsymbol{\Omega}',s')\Psi(\boldsymbol{x},\boldsymbol{\Omega}',s')ds'd\Omega' + Q(\boldsymbol{x})  \right], \quad \boldsymbol{x} \in V, \:\boldsymbol{\Omega}\in 4\pi, \: 0<s.
\end{aligned}\vspace*{-0.1cm}
\end{flalign}
The notation is standard \cite{Larsen01}: $\boldsymbol{x} = (x,y,z)$ is a point of space, $\boldsymbol{\Omega} = (\Omega_{x},\Omega_{y},\Omega_{z})$ is the particle's direction of flight and $s$ is the path-length, i.e., the distance traveled by the particle since its last interaction (birth or scattering). Furthermore, $\Psi$ is the nonclassical angular flux, $Q$ is a source, $c$ is the scattering ratio and $\Sigma_{t}$ is the angular-dependent total macroscopic cross section, which satisfies the relation\vspace*{-0.3cm}
\begin{equation}\label{e1.1b}
p(\boldsymbol{\Omega},s) = \Sigma_{t}(\boldsymbol{\Omega},s)e^{-\int_{0}^{s}\Sigma_{t}(\boldsymbol{\Omega},s')ds'},
\end{equation}
where $p(\boldsymbol{\Omega},s)$ is the free-path distribution function.
\label[pluraleq]{e1.1}
\end{subequations}

The nonclassical transport theory has received increasing attention in the last few years, being applied in  a number of different areas, including nuclear engineering \cite{Vasques15} and computer graphics \cite{Bitterli18}. Recently, in the work \cite{Leonardo20}, a mathematical approach that allows one to solve \cref{e1.1a} in a deterministic fashion was first described. In this approach, namely Spectral Approach (SA), we consider $\Psi$ as
\begin{subequations}
	\begin{equation}\label{e1.2a}
	\Psi(\boldsymbol{x},\boldsymbol{\Omega},s) =\psi(\boldsymbol{x},\boldsymbol{\Omega},s) e^{-\int_{0}^{s}\Sigma_{t}(\boldsymbol{\Omega},s')ds'}.
	\end{equation}
	Here $\psi$ is represented as a truncated series of Laguerre polynomials in $s$. That is,
	\begin{equation}\label{e1.2b}
\psi(\boldsymbol{x},\boldsymbol{\Omega},s) = \sum_{m=0}^{M}\psi_{_{m}}(\boldsymbol{x},\boldsymbol{\Omega})L_{_{m}}(s),
	\end{equation}
where $M$ is the truncation order for the Laguerre series and $L_{_{m}}(s)$ is the Laguerre polynomial of degree $m$. 
The function $\psi_{_{m}}(\boldsymbol{x},\boldsymbol{\Omega})$ is the solution of the equation \cite{Leonardo20}
	\small{\begin{flalign}\label{e1.2c}
	&\boldsymbol{\Omega}\boldsymbol{\cdot}\nabla\psi_{_{m}}(\boldsymbol{x},\boldsymbol{\Omega}) + \sum_{j=0}^{m}\psi_{j}(\boldsymbol{x},\boldsymbol{\Omega}) = \frac{1}{4\pi}\left[c \int_{4\pi}^{}\sum_{k=0}^{M}\psi_{_{k}}(\boldsymbol{x},\boldsymbol{\Omega}')\mathcal{L}_{_{k}}(\boldsymbol{\Omega}')d\Omega' + Q(\boldsymbol{x})\right], m = 0,1,\dots,M,
	\end{flalign}} \normalsize
	where we have used the definition
	\begin{equation}\label{e1.2d}
	\mathcal{L}_{_{k}}(\boldsymbol{\Omega}) = \int_{0}^{\infty}p(\boldsymbol{\Omega},s)L_{_{k}}(s)ds.
	\end{equation} 

	At this point we remark that, for some problems, the truncation order $M$, required to generate accurate results for $\Psi$, is substantially large. This situation increases considerably the computer running-timing spent to estimate $\Psi$. Moreover, the increase of the truncation order with the hope of obtaining more accurate results may have the opposite effect, depending the method one uses to solve \cref{e1.2}. This situation is characterized in reference \cite{Moraes2021Jun}, where the authors used an analytic nodal method to solve \cref{e1.2}. In fact, accuracy of the solution was negatively affected, in some cases, due to computational finite precision arithmetic.
	
 Bearing this in mind, we describe in \cref{sec2} a simple and useful modification of the Spectral Approach \cite{Leonardo20}. In the present modified SA we slightly change the representation of $\Psi$ (\cref{e1.2a}) which, for some cases, decreases considerably the truncation order for the Laguerre series required to generate accurate results. It is also demonstrated in \cref{sec2} that the standard SA is a particular case of the modified SA. In \cref{sec3}, we present a practical application where the modified SA decreases the truncation order of the Laguerre series, in comparison with the standard SA, generating results with similar accuracy. We conclude with a brief discussion in \cref{sec4}.
	\label[pluraleq]{e1.2}
\end{subequations} 

\section{The modified spectral approach}\label{sec2}
\setcounter{equation}{0}
The idea behind the modified SA is to change the shape of functions $\mathcal{L}_{_{k}}(\boldsymbol{\Omega})$, making them go to zero quickly with the increase of $k$. Thus, it is expected, for some cases, that the truncation order $M$ for the Laguerre series, required to generate accurate results, decrease.

Let us write \cref{e1.1a} in its $``$initial value'' form \cite{Leonardo20}:
\begin{subequations}
	\begin{equation}\label{e2.1a}
	\frac{\partial}{\partial s}\Psi(\boldsymbol{x},\boldsymbol{\Omega},s) + \boldsymbol{\Omega}\boldsymbol{\cdot}\nabla\Psi(\boldsymbol{x},\boldsymbol{\Omega},s) + \Sigma_{_{t}}(\boldsymbol{\Omega},s)\Psi(\boldsymbol{x},\boldsymbol{\Omega},s) = 0,
	\end{equation}\vspace*{-0.7cm}
	\begin{equation}\label{e2.1b}
	\Psi(\boldsymbol{x},\boldsymbol{\Omega},0) = \frac{1}{4\pi}\left[c\int_{4\pi}^{}\int_{0}^{\infty}\Sigma_{_{t}}(\boldsymbol{\Omega}',s')\Psi(\boldsymbol{x},\boldsymbol{\Omega}',s')ds'd\Omega' + Q(\boldsymbol{x})  \right].
	\end{equation}
We define $\Psi$ slightly different from \cref{e1.2a}. That is,
	\label[pluraleq]{e2.1}
\end{subequations}
\begin{subequations}
	\begin{equation}\label{e2.2a}
\Psi(\boldsymbol{x},\boldsymbol{\Omega},s) = \widehat{\psi}(\boldsymbol{x},\boldsymbol{\Omega},s)e^{-\int_{0}^{s}\Upsilon(\boldsymbol{\Omega},s')ds'},
	\end{equation}
	where 
		\begin{equation}\label{e2.2b}
\Upsilon(\boldsymbol{\Omega},s) = \alpha+\Sigma_{_{t}}(\boldsymbol{\Omega},s),
	\end{equation}
	with $\alpha$ being an arbitrary constant. Substituting \cref{e2.2} into \cref{e2.1} we obtain
		\label[pluraleq]{e2.2}
\end{subequations}
\begin{subequations}
	\begin{equation}\label{e2.3a}
\frac{\partial}{\partial s}\widehat{\psi}(\boldsymbol{x},\boldsymbol{\Omega},s) + \boldsymbol{\Omega}\boldsymbol{\cdot}\nabla\widehat{\psi}(\boldsymbol{x},\boldsymbol{\Omega},s) -\alpha\,\widehat{\psi}(\boldsymbol{x},\boldsymbol{\Omega},s) = 0,
\end{equation}\vspace*{-0.7cm}
\begin{equation}\label{e2.3b}
\widehat{\psi}(\boldsymbol{x},\boldsymbol{\Omega},0) = \frac{1}{4\pi}\left[c\int_{4\pi}^{}\int_{0}^{\infty}e^{-\alpha s'}p(\boldsymbol{\Omega}',s')\widehat{\psi}(\boldsymbol{x},\boldsymbol{\Omega}',s')ds'd\Omega' + Q(\boldsymbol{x})  \right].
\end{equation}
To proceed, we represent $\widehat{\psi}$ as a truncated series of Laguerre polynomials in $s$, i.e.,
		\label[pluraleq]{e2.3}
\end{subequations}
\begin{equation}\label{e2.4}
\widehat{\psi}(\boldsymbol{x},\boldsymbol{\Omega},s) = \sum_{m=0}^{M}\widehat{\psi}_{_{m}}(\boldsymbol{x},\boldsymbol{\Omega})L_{_{m}}(s),
\end{equation}
which we substitute into \cref{e2.3}, thus obtaining
\begin{subequations}
	\begin{equation}\label{e2.5a}
	\sum_{m=0}^{M}\widehat{\psi}_{_{m}}(\boldsymbol{x},\boldsymbol{\Omega})\frac{d}{ds}L_{_{m}}(s) + \sum_{m=0}^{M}\boldsymbol{\Omega}\boldsymbol{\cdot}\nabla\widehat{\psi}_{_{m}}(\boldsymbol{x},\boldsymbol{\Omega})L_{_{m}}(s) - \alpha\,\sum_{m=0}^{M}\widehat{\psi}_{_{m}}(\boldsymbol{x},\boldsymbol{\Omega})L_{_{m}}(s) = 0,
		\end{equation}
	and
		\vspace*{-0.5cm}
		\begin{equation}\label{e2.5b}
		\sum_{j = m+1}^{M}\widehat{\psi}_{_{j}}(\boldsymbol{x},\boldsymbol{\Omega}) = \frac{1}{4\pi}\left[c\int_{4\pi}^{} \sum_{k=0}^{M}\widehat{\psi}_{_{k}}(\boldsymbol{x},\boldsymbol{\Omega})\widehat{\mathcal{L}}_{_{k}}(\boldsymbol{\Omega}')d\Omega' + Q(\boldsymbol{x})\right] - \sum_{j=0}^{m}\widehat{\psi}_{_{j}}(\boldsymbol{x},\boldsymbol{\Omega}),
		\end{equation}
		where 
		\begin{equation}\label{e2.5c}
	\widehat{\mathcal{L}}_{_{k}}(\boldsymbol{\Omega}) = \int_{0}^{\infty}e^{-\alpha s}p(\boldsymbol{\Omega},s)L_{_{k}}(s)ds.
		\end{equation}
		\label[pluraleq]{e2.5}
		Multiplying \cref{e2.5a} by $e^{-s}L_{_{n}}(s)$ and operating the resulting equation by $\int_{0}^{\infty}(\cdot)ds$, we obtain\vspace*{-0.3cm}
\end{subequations}
\begin{equation}\label{e2.6}
 \boldsymbol{\Omega}\boldsymbol{\cdot}\nabla\widehat{\psi}_{_{m}}(\boldsymbol{x},\boldsymbol{\Omega}) -\alpha\,\widehat{\psi}_{_{m}}(\boldsymbol{x},\boldsymbol{\Omega}) = \sum_{j=m+1}^{M}\widehat{\psi}_{_{j}}(\boldsymbol{x},\boldsymbol{\Omega}).
\end{equation}
Substituting \cref{e2.5b} into \cref{e2.6} yields
\begin{equation}\label{e2.7}
 \boldsymbol{\Omega}\boldsymbol{\cdot}\nabla\widehat{\psi}_{_{m}}(\boldsymbol{x},\boldsymbol{\Omega})  + \sum_{j=0}^{m}\widehat{\psi}_{_{j}}(\boldsymbol{x},\boldsymbol{\Omega})      -\alpha \,\widehat{\psi}_{_{m}}(\boldsymbol{x},\boldsymbol{\Omega}) =  \frac{1}{4\pi}\left[c\int_{4\pi}^{} \sum_{k=0}^{M}\widehat{\psi}_{_{k}}(\boldsymbol{x},\boldsymbol{\Omega})\widehat{\mathcal{L}}_{_{k}}(\boldsymbol{\Omega}')d\Omega' + Q(\boldsymbol{x})\right].
\end{equation}

For the particular case where $\alpha=0$, the modified SA is equivalent to the standard SA, since \cref{e2.2a,e2.5c,e2.7} become identical to Eqs. (\ref{e1.2a}), (\ref{e1.2d}) and (\ref{e1.2c}) respectively.

\section{A practical example}\label{sec3}
\setcounter{equation}{0}
To begin, we consider \cref{e2.7} for slab-geometry problems in the discrete ordinates (S$_{_{\text{N}}}$) formulation \cite{HB10}. That is,
\begin{equation}\label{e3.1}
\mu_{_{n}}\frac{d}{dx}\widehat{\psi}_{_{m,n}}(x) + \sum_{j=0}^{m}\widehat{\psi}_{_{j,n}}(x)      -\alpha \,\widehat{\psi}_{_{m,n}}(x) =  \frac{1}{2}\left[c \sum_{k=0}^{M}\sum_{\ell=1}^{N}\widehat{\psi}_{_{k,\ell}}(x)\widehat{\mathcal{L}}_{_{k,\ell}}\omega_{\ell} + Q(x)\right],
\end{equation}
where $\mu_{_{n}}$ is a discrete direction of motion, such that $n = 1,2,\dots,N$, with $N$ being the order of the angular quadrature with weights $\omega_{n}$. The subscript that appears in $\widehat{\psi}_{_{m,n}}$ and $\widehat{\mathcal{L}}_{_{m,n}}$ indicates that these functions are evaluated with respect to the discrete directions. In other words, $\widehat{\psi}_{_{m}}(x,\mu_{_{n}}) = \widehat{\psi}_{_{m,n}}(x)$ and $\widehat{\mathcal{L}}_{_{m}}(\mu_{_{n}}) = \widehat{\mathcal{L}}_{_{m,n}}$.

To present a practical example of potential benefits of the modified SA, we consider reference \cite{Moraes2021Jun} wherein the authors seek to analyze the application of the Analytical Discrete Ordinates (ADO) method \cite{Lili95} to the solution of \cref{e1.2c}. In that paper, the authors reproduced the solution of the classical diffusion equation through the nonclassical relation
\begin{subequations}
\begin{equation}\label{e3.2a}
\Phi(x) = \frac{1}{\sigma_{_{t}}}\sum_{m=0}^{M}\widehat{\mathcal{L}}_{_{m}}\sum_{n=1}^{N} \widehat{\psi}_{_{m,n}}(x) \omega_{_{n}},
\end{equation}
 where $\Phi$ is the diffusion scalar flux and $\sigma_{_{t}}$ is the macroscopic total cross section considered in the diffusion problem. In \cref{e3.2a}, $\widehat{\psi}_{_{m,n}}$ is the solution of \cref{e3.1}, considering the following free-path distribution function
 \begin{equation}\label{e3.2b}
 p_{_{m}}(\mu_{_{n}})= 3\, \sigma^{2}_{_{t}} se^{^{-\sqrt{3}\,\sigma_{_{t}}s}}
 \end{equation}
in the calculation of functions $\widehat{\mathcal{L}}_{_{m}}$. We remark that functions $\widehat{\mathcal{L}}_{_{m}}$ are, in this case, independent of the discrete direction of motion due to the considered free-path distribution function. More details about the relation between the nonclassical equation and the classical diffusion equation can be found in reference \cite{Moraes2021Jun}
 \label[pluraleq]{e3.2}
\end{subequations}

To proceed, we focus our attention on the second model problem of reference \cite{Moraes2021Jun}. This model problem has the spatial domain $\mathcal{D} = \left\lbrace x \in \mathbb{R} \left| \right. 0\leq x \leq 20 \right\rbrace$, with an isotropic source 
\begin{equation}\label{e3.3}
Q(x) = \left\lbrace \begin{array}{cc}
1, & 9.5 \leq x \leq 10.5 \\
0, & \text{otherwise}
\end{array}\right..
\end{equation}
In addition, the total macroscopic cross section is $\sigma_{_{t}} = 1$ and the scattering ratio is $c = 0.5$. To completely establish this problem we consider vacuum boundary conditions, i.e.,
	\begin{equation}\label{e3.3a}
\widehat{\psi}_{_{m,n}}(x^{\star}) = 0,\text{ with } x^{\star} = \left\lbrace \begin{array}{cc}
	0, & \text{if }\mu_{_{n}}>0 \\
	20, & \text{if }\mu_{_{n}}<0
\end{array}\right..
	\end{equation}
 
\begin{figure}
 	\centering
 	\begin{threeparttable}
 		\begin{tabular}{c}
 				\includegraphics[scale=0.63]{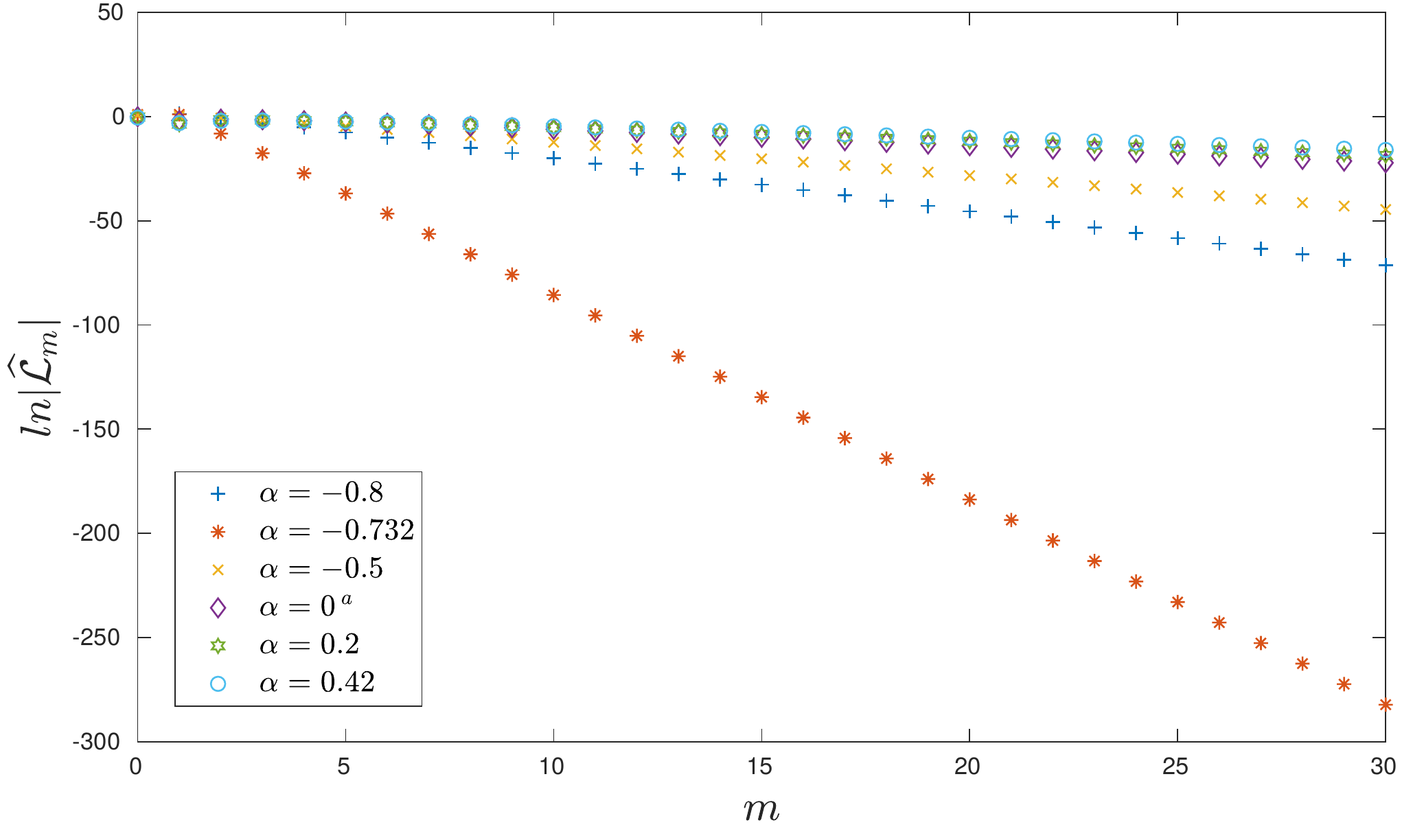}
 		\end{tabular}
 		\vspace*{-0.3cm}
 		\begin{tablenotes}[]
 			\tiny 
 			\item[$\qquad \qquad \:\: a$]  Standard spectral approach.
 		\end{tablenotes}
 	\end{threeparttable}
 	\caption{Functions $\widehat{\mathcal{L}}_{_{m}}$ defined in \cref{e2.5c}.}
 	\label{f1}
 \end{figure}

 Given the total cross section considered in this example and the free-path distribution function as in \cref{e3.2b}, we calculate functions $\widehat{\mathcal{L}}_{_{m}}$ (\cref{e2.5c}) for different values of $\alpha$, and depicted some of these results in \cref{f1}. 
As we can notice in \cref{f1}, functions $\widehat{\mathcal{L}}_{_{m}}$ go to zero very quickly when $\alpha = -0.732$. In fact, for $\alpha = 0$ (standard SA), the functions $\widehat{\mathcal{L}}_{_{m}}$ go to zero in a much slower fashion in comparison to $\alpha = -0.732$. Therefore, we expect to achieve better results using $\alpha = -0.732$ than using $\alpha = 0$, for a fixed truncation order $M$ of the Laguerre series. 

In order to verify our assertion, we used the ADO method, described in reference \cite{Moraes2021Jun}, to solve \cref{e3.1}
 with $N=30$ (Gauss-Legendre angular quadrature \cite{HB10}), considering $\alpha = -0.732$ and $\alpha = 0$ for a number of different truncation orders. Then, we use these solutions to reproduce the diffusion scalar fluxes according to \cref{e3.2a}.
\begin{table}
	\caption{Diffusion scalar flux.}
	\vspace*{-0.5cm}
	\small
	\begin{center}
		{\renewcommand{\arraystretch}{1.15}
			\adjustbox{angle=00}{
				\begin{threeparttable}
				%	\resizebox{5cm}{!}{
\begin{tabular}{ccccc|ccc}
	\toprule 
	\multirow{3}{*}{$x^{\star\tnote{a}}$} & \multirow{3}{*}{Benchmark\tnote{b}} & \multicolumn{3}{c}{Nonclassical solution (\cref{e3.2a})} & \multicolumn{3}{|c}{Relative deviation}\tabularnewline
	\cmidrule{3-8} \cmidrule{4-8} \cmidrule{5-8} \cmidrule{6-8} \cmidrule{7-8} \cmidrule{8-8} 
	&  & \multicolumn{6}{c}{$\alpha = 0$}\tabularnewline
	\cmidrule{3-8} \cmidrule{4-8} \cmidrule{5-8} \cmidrule{6-8} \cmidrule{7-8} \cmidrule{8-8} 
 &  & $M = 2$ & $M = 10$ & $M = 20$ & $M = 2$ & $M = 10$ & $M = 20$\tabularnewline
\midrule 
0.0 & 1.83175E+00\tnote{c} & 1.851197E+00 & 1.831641E+00 & 1.83172E+00 & 1.06E-02 & 5.75E-05 & 1.18E-05\tabularnewline
\midrule 
2.0 & 2.24951E-01 & 2.458183E-01 & 2.249163E-01 & 2.24951E-01 & 9.28E-02 & 1.53E-04 & 2.75E-07\tabularnewline
\midrule 
4.0 & 1.94217E-02 & 2.409218E-02 & 1.942560E-02 & 1.94217E-02 & 2.40E-01 & 2.00E-04 & 3.90E-07\tabularnewline
\midrule 
6.0 & 1.67681E-03 & 2.910856E-04 & 1.674713E-03 & 1.67681E-03 & 8.26E-01 & 1.25E-03 & 3.08E-07\tabularnewline
\midrule 
8.0 & 1.44588E-04 & -6.597009E-04 & 1.456960E-04 & 1.44588E-04 & 5.56E+00 & 7.66E-03 & 2.85E-07\tabularnewline
\midrule 
10.0 & 1.03548E-05 & -1.793719E-04 & 1.023733E-05 & 1.03564E-05 & 1.83E+01 & 1.13E-02 & 1.53E-04\tabularnewline
\midrule 
&  & \multicolumn{6}{c}{$\alpha = -0.732$}\tabularnewline
\midrule 
&  & $M = 0$ & $M = 1$ & $M = 2$ & $M = 0$ & $M = 1$ & $M = 2$\tabularnewline
\midrule 
0.0 & 1.83175E+00 & 1.17988E+01 & 1.83192E+00 & 1.83172E+00 & 5.44E+00 & 9.53E-05 & 1.19E-05\tabularnewline
\midrule 
2.0 & 2.24951E-01 & 1.56834E+00 & 2.24957E-01 & 2.24951E-01 & 5.97E+00 & 2.58E-05 & 2.44E-07\tabularnewline
\midrule 
4.0 & 1.94217E-02 & 1.94635E-01 & 1.94227E-02 & 1.94217E-02 & 9.02E+00 & 5.23E-05 & 2.76E-07\tabularnewline
\midrule 
6.0 & 1.67681E-03 & 2.43820E-02 & 1.67695E-03 & 1.67681E-03 & 1.35E+01 & 8.14E-05 & 5.14E-07\tabularnewline
\midrule 
8.0 & 1.44588E-04 & 3.03975E-03 & 1.44604E-04 & 1.44588E-04 & 2.00E+01 & 1.10E-04 & 3.04E-07\tabularnewline
\midrule 
10.0 & 1.03548E-05 & 1.96605E-04 & 1.03577E-05 & 1.03564E-05 & 1.80E+01 & 2.80E-04 & 1.57E-04\tabularnewline
	\bottomrule
\end{tabular}%}
					\begin{tablenotes}[flushleft]
						\tiny
						\item[a] $x = \pm x^{\star}+10.0$. For example, if $x^{\star} = 2.0$, the results presented are valid for $x = 8.0$ and $x = 12.0$. We use $x^{\star}$ due to the problem's symmetry at $x = 10.0$.
						\item[b] Table 7 of reference \cite{Moraes2021Jun}.
						\item[c] Read as 1.83175$\times10^{+00}$.
					\end{tablenotes}
				\end{threeparttable}
			}
		}
	\end{center}
	\label{t1}
\end{table}
Analyzing \cref{t1}, we observe that for the same truncation order $M=2$ for the Laguerre series, the solution with $\alpha = -0.732$ generated a maximum relative deviation of $2.80\times 10^{^{-4}}$ compared to $1.83$ produced with $\alpha = 0$. In fact, for the standard SA ($\alpha$ = 0), considering $M=2$, even negative values were generated. Moreover, the standard SA produced a solution with similar accuracy as that for $\alpha = -0.732$ ($M=2$), only considering $M=20$.  Therefore, using the modified SA (with $\alpha = -0.732$) we were able to decrease the truncation order for the Laguerre series from $M=20$ to $M=2$, in this case.

 \vspace*{-0.2cm}
 \section{Discussion}\label{sec4}
 \setcounter{equation}{0}
 In this work we described a simple and useful modification of the spectral approach \cite{Leonardo20} used for solving nonclassical transport problems in a deterministic fashion. This modification, in some cases, allows the decrease of the truncation order $M$ of the Laguerre series used to represent the nonclassical angular flux, as we presented in the previous section. 
 
 At this point we remark that a good choice of $\alpha$ is important for the efficiency of the modified SA. However, the best choice of $\alpha$ is not always simple to be found  and depends on the free-path distribution function considered. In addition, the use of $\alpha \neq 0$ can lead to numerical concerns due to the exponential $e^{-\alpha s}$, which appears in \cref{e2.5c}.
 
 Further work needs to be done in order to analyze convergence of the modified SA. This analysis will also include the convergence analysis of the standard Spectral Approach, since it is a particular case of the modified SA.
 
 \vspace*{-0.2cm}
 \small{
\section*{Acknowledgments}
This study was financed in part by the Coordena\c c\~ao de Aperfei\c coamento de Pessoal de N\'ivel Superior - Brasil (CAPES) - Finance Code 001, and Funda\c c\~ao Carlos Chagas Filho de Amparo \`a Pesquisa do Estado do Rio de Janeiro - Brasil (FAPERJ).
L.R.C.~Moraes and R.C.~Barros also acknowledge support from Conselho Nacional de Desenvolvimento Cient\'ifico e Tecnol\'ogico - Brasil (CNPq).
J. K.~Patel and R.~Vasques acknowledge support under award number NRC-HQ-84-15-G-0024 from the Nuclear Regulatory Commission.
}
\vspace*{-0.2cm}
\small{
\bibliography{Notas}
}
\end{document}